\documentclass[journal]{IEEEtran}
\usepackage{graphicx}

\begin{document}
\title{Characterization of GEM Detectors for Application in the
  CMS Muon Detection System\\
\vspace*{-4cm}{\tiny This work has been submitted to the IEEE
  Nucl. Sci. Symp. 2010 for publication in the conference record. Copyright
  may be transferred without notice, after which this version may no longer be
  available.}\\
\hspace*{15cm}{\bf\small RD51-Note-2010-005}\vspace*{2cm}}
%
%

\author{D.~Abbaneo,
        S.~Bally,
        H.~Postema,
        A.~Conde~Garcia,
        J.~P.~Chatelain,
        G.~Faber,
        L.~Ropelewski,
        E.~David,
        S.~Duarte~Pinto,
        G.~Croci,
        M.~Alfonsi,
        M.~van~Stenis,
        A.~Sharma,~\IEEEmembership{Senior Member,~IEEE,}
        L.~Benussi,
        S.~Bianco,
        S.~Colafranceschi,
        D.~Piccolo,
        G.~Saviano,
        N.~Turini,
        E.~Oliveri,
        G.~Magazzu',
        A.~Marinov,
        M.~Tytgat*,~\IEEEmembership{Member,~IEEE,}
        N.~Zaganidis,
        M.~Hohlmann,~\IEEEmembership{Member,~IEEE,}
        K.~Gnanvo,
        Y.~Ban,
        H.~Teng,
        J.~Cai

\thanks{Manuscript received November 19, 2010}
\thanks{D.~Abbaneo, S.~Bally, H.~Postema, A.~Conde~Garcia, J.-P.~Chatelain,
  G.~Faber, L.~Ropelewski, E.~David, S.~Duarte~Pinto, G.~Croci, M.~Alfonsi, M.~van~Stenis, A.~Sharma,
       are with CERN, Geneva, Switzerland}
\thanks{S.~Colafranceschi is with CERN, Geneva, Switzerland and Laboratori Nazionali di Frascati dell'INFN, Frascati, Italy and Sapienza Universit\`a di Roma - Facolta' Ingegneria}
\thanks{L.~Benussi, S.~Bianco, D. Piccolo are with Laboratori Nazionali di Frascati dell'INFN, Frascati, Italy}
\thanks{G.~Saviano is with Sapienza Universit\`a di Roma - Facolta' Ingegneria, Rome, Italy and Laboratori Nazionali di Frascati dell'INFN, Frascati, Italy}
\thanks{N.~Turini, E.~Oliveri, G.~Magazzu' are with INFN, Sezione di Pisa, Universit\`a Degli Studi di Siena, Siena, Italy}
\thanks{A.~Marinov, M.~Tytgat, N.~Zaganidis are with the Department of
  Physics and Astronomy, Universiteit Gent, Gent, Belgium}
\thanks{M.~Hohlmann, K.~Gnanvo are with Dept. of Physics and Space Sciences, Florida Institute of Technology, Melbourne, FL, USA}
\thanks{Y.~Ban, H.~Teng, J.~Cai are with Peking University, Beijing, China}
\thanks{* Corresponding author, michael.tytgat@cern.ch}
}

\maketitle
\pagestyle{empty}
\thispagestyle{empty}

\begin{abstract}
The muon detection system of the Compact Muon Solenoid 
experiment at the CERN Large Hadron Collider is based on
different technologies for muon tracking and triggering. In particular, the
muon system in the endcap disks of the detector consists of Resistive Plate Chambers for
triggering and Cathode Strip Chambers for tracking. At present, the endcap
muon system is only partially instrumented with the very forward detector region
remaining uncovered.  In view of a possible future extension of the
muon endcap system, we report on a feasibility study on the use of
Micro-Pattern Gas Detectors, in particular Gas Electron Multipliers, for both muon
triggering and tracking. Results on the construction and characterization of small
triple-Gas Electron Multiplier prototype detectors are presented.
\end{abstract}


\section{Introduction}
\begin{figure}[!t]
\centering
\includegraphics[width=3.5in]{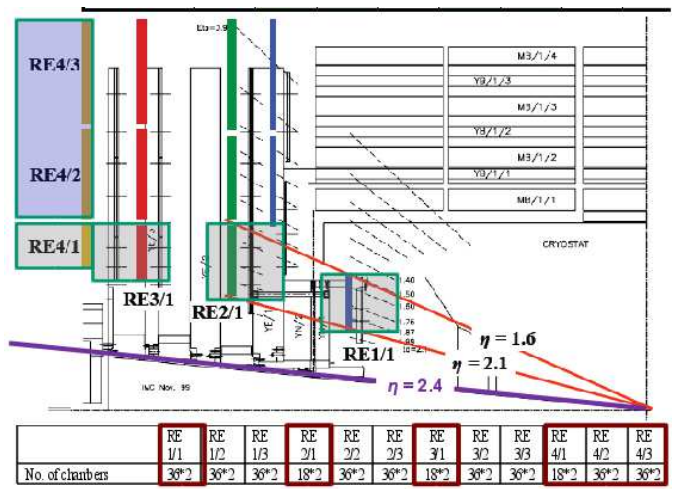}
\caption{The CMS RPC endcap system. RPCs (REi/1 and RE4/2-3) enclosed in boxes
  have yet 
  to be installed.} 
\label{fig:CMSendcapdiag}
\end{figure}

%
%
%
%
\IEEEPARstart{T}{he} muon system of the Compact Muon Solenoid (CMS) 
detector~\cite{cmsmuondetector} at
the CERN Large Hadron Collider (LHC)  
is based on
different technologies for muon tracking and triggering. 
While Drift
Tubes and Cathode Strip Chambers provide muon tracking in the barrel and
endcap region, respectively, Resistive Plate Chambers (RPCs) are used for 
level 1 muon triggering in both the barrel and endcap detector parts.
The latter RPC 
endcap chambers are double-gap
Bakelite based RPCs with strip readout.  
They are operated in avalanche mode
at 9.5~kV with a 
C$_2$H$_2$F$_4$:iC$_4$H$_{10}$:SF$_6$ 
(94.7:5.0:0.3) gas mixture humidified at about 40\%.
As shown in Figure~\ref{fig:CMSendcapdiag}, the RPC endcap  
system 
is presently incomplete as only the low pseudo-rapidity region, $|\eta| <
1.6$, of the three existing endcap disks is equipped with chambers, REi/2-3. 
To maintain a good muon trigger performance during the future 
LHC running at full luminosity and beam energy,
an extension of the RPC endcap system is foreseen for 2012 with the addition
of a fourth station with RE4/2-3 chambers. 
However, the
very forward region, $|\eta|>1.6$,  of all endcap disks will remain
uninstrumented and presents an
opportunity to add more capable and robust detectors in the vacant REi/1 locations.
For the planned LHC luminosity upgrade ($\sim10^{34-35}$~cm$^2$/s), the expected
particle flux in that detector region amounts to several tens
of kHz with a total integrated charge over 10 years of several tens of
$\mbox{C}/\mbox{cm}^2$.  To cope with such a hostile environment, technologies
other than the existing Bakelite RPCs will have to be found.

\section{The Case for Micro-Pattern Gas Detectors}
 
A dedicated R\&D program was launced in 2009 to study the feasibility of using
micro-pattern gas detectors (MPGD) for the instrumentation of the vacant 
$|\eta|>1.6$ region in the present RPC endcap system.
Micro-pattern gas detectors can offer an excellent spatial resolution of order
100~$\mu$m, a time resolution below 5~ns, a good overall detector
efficiency above 98\% and
a rate capability of order $10^6$~Hz/mm$^2$ that is sufficient to handle the 
expected particle
fluxes in the LHC environment. 
In the case of the existing RPC system, the large volume, the cost of the gas
mixture, 
and the
need to constantly remove impurities from the gas circuit to guarantee a 
stable detector
operation, make the use of a rather complex closed-loop gas system including
filtering mandatory.
For MPGDs, their operation with a non-flammable gas mixture, e.g. Ar:CO$_2$, is
therefore also advantageous compared to the present RPC system.  

With the enhanced $(\eta,\phi)$ readout granularity and rate
capability of the MPGDs, one could effectively improve the level 1 muon
trigger 
efficiency and even offer both triggering and tracking functions at the same time. 
In this case, one could
even consider to extend the pseudo-rapidity range of the system up to
$|\eta|=2.4$,
to match the coverage of the Cathode Strip Chambers in the endcaps.

\section{Small MPGD Prototypes}

\begin{figure}[!t]
\centering
\includegraphics[width=2.5in]{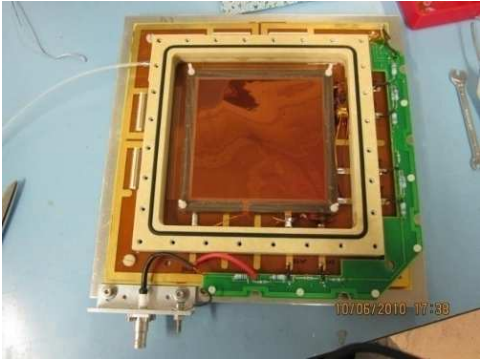}\vspace*{.5cm}
\includegraphics[width=2.5in]{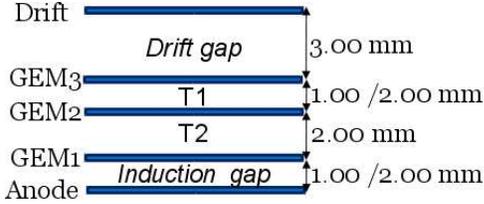}
\caption{The standard double-mask triple-GEM prototype. The picture on top
 shows the detector during assembly; the diagram at the bottom depicts the gap
sizes.}
\label{fig:standardgem}
\end{figure}

\begin{figure}[!t]
\centering
\includegraphics[width=2.5in]{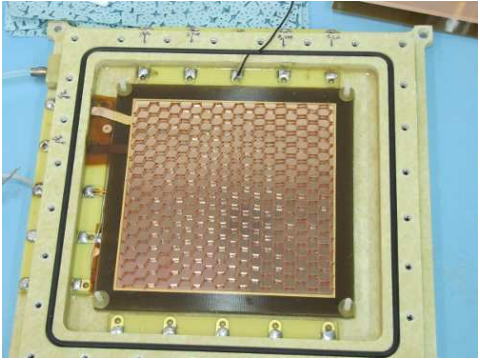}
\caption{The triple-GEM prototype with honeycomb spacers.}
\label{fig:honeycombgem}
\end{figure}

In a first step of the study a characterization was done of two different small
MPGD prototypes~: one Micromegas~\cite{micromegas} and one 
triple-GEM~\cite{gem} detector. 
Both prototypes with an active area of $10\times 10$~cm$^2$ were produced in
the CERN EN-ICE surface treatment workshop and were 
subsequently tested in the RD51~\cite{rd51} lab of the CERN Detector
Technology 
Group (DT).  
Using standard Ar:CO$_2$ gas mixtures, the two detectors were characterized by 
measuring gain and
pulse height spectra with radioactive sources and Cu X-rays from a generator.
Their efficiency plateaus were measured and the optimal 
operational 
voltages were determined. 
In October 2009, the two prototypes were put into a pion/muon
test beam at the CERN SPS H4 beam line~\cite{SPSH4beamline}.
A good detector performance was observed for 
the 
triple-GEM, while the Micromegas prototype showed
a substantial discharge probability and hence a poor data quality.
The discharge probabilities of the two detectors were measured in the RD51
lab. For the triple-GEM a probability of $10^{-6}$ was measured for gains up to
$2\cdot 10^4$, while the Micromegas was discharging with a probability of
$10^{-4}$ at a gain of less than 2000, which is consistent with previous
studies~\cite{SauliArchana}. Consequently, based on these findings and the available
general expertise on GEMs in the research group, the triple-GEM prototype 
was selected for further studies. 

The triple-GEM mentioned above 
was constructed using the standard double-mask technique for the etching
of the GEM foils. The foils are made of 50~$\mu$m thick kapton sheets with
a 5~$\mu$m copper cladding on both sides. The GEM and cathode drift foils were
glued on fiberglass frames and mounted inside a gas-tight box as shown in 
Figure~\ref{fig:standardgem}. The detector 
has 128 strips with
a pitch of
0.8~mm. Two different gap size configurations were tried to study the effect
on the detector performance (drift, transfer 1, transfer 2,
induction gap size): 3/2/2/2~mm and 3/1/2/1~mm.

In addition to the standard double-mask GEM prototype another 10$\times$10~cm$^2$ 
triple-GEM
prototype was
constructed using the single-mask technique~\cite{singlemask}, which overcomes
the problems with the alignment of the masks on either side of the foils
during the photolitographic etching of the holes. This prototype has 256
strips in two perpendicular directions, with a strip pitch of 0.4~mm.
 
Futhermore, to avoid the need for foil stretching during detector assembly, 
a technique based on inserting honeycomb spacers into the detector gaps was
tested with a triple-GEM as shown in Figure~\ref{fig:honeycombgem}. Three
different configurations were tried with varying honeycomb cell sizes (in drift,
transfer 1, transfer 2, induction gap): 12/12/12/12~mm (``config 1''),
6/12/12/12~mm (``config 2''), 6/0/0/0~mm (``config 3'').
This prototype also has 256
strips in two perpendicular directions, with a strip pitch of 0.4~mm. 
 
\section{Test Beam Measurements}

\subsection{Setup}

\begin{figure}[!t]
\centering
\includegraphics[width=3.5in]{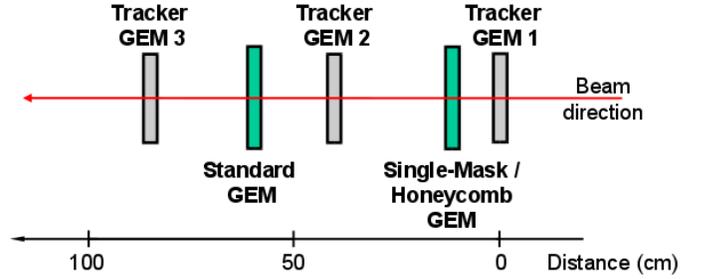}
\caption{The RD51 beam telescope at the CERN SPS H4 beam line. The
  triple-GEMs of the telescope are labeled ``Tracker GEM''.}
\label{fig:rd51beamtelescope}
\end{figure}

\begin{figure}[!t]
\centering
\includegraphics[width=3.5in]{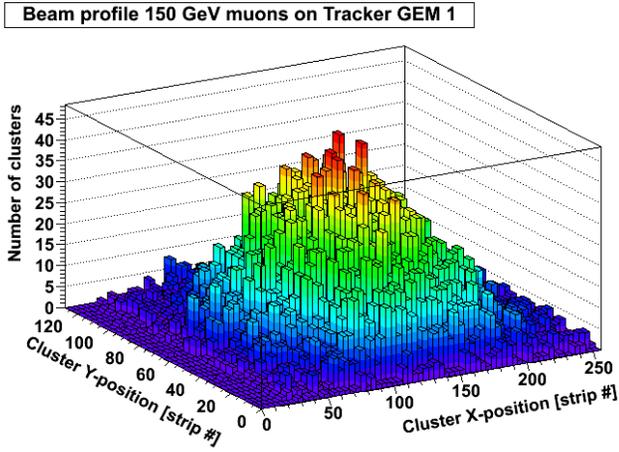}\vspace*{.5cm}
\includegraphics[width=3.5in]{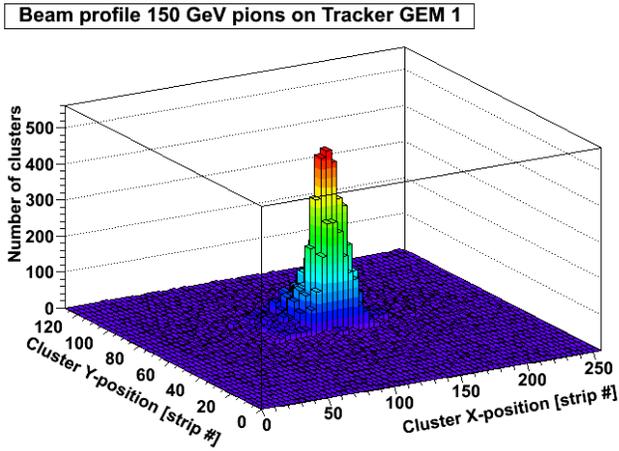}
\caption{The beam profiles for muons and
  pions obtained with the tracker GEMs.}
\label{fig:beamprofiles}
\end{figure}

The triple-GEM 
prototypes were tested with a 150 GeV muon/pion beam at the CERN SPS
H4 beam line during several RD51 test beam campaigns. 
The detectors under test were mounted into
the RD51 triple-GEM beam telescope as depicted in
Figure~\ref{fig:rd51beamtelescope}. The telescope consists of three standard
triple-GEM detectors, hereafter referred to as tracker GEMs, 
with a 10$\times$10~cm$^2$ active area, running with a
Ar:CO$_2$ (70:30) gas mixture. They have 256 strips in both horizontal
(y-coordinate) 
and 
vertical (x-coordinate) directions
transverse to the beam, with a pitch of 0.4~mm. The telescope detectors were always
operated at a gain larger than 10$^4$. This setup served as tracking
device for the detectors under test.

The standard double-mask triple-GEM prototypes under test were studied with 
different gas
mixtures, Ar:CO$_2$ (70:30, 90:10) and Ar:CO$_2$:CF$_4$
(45:15:40, 60:20:20), with a gas flow of about 5~l/hour corresponding to
roughly 50 detector volume exchanges per hour.
The single-mask and
honeycomb triple-GEMs were operated with an Ar:CO$_2$ mixture only.

The readout of all detectors including the tracker GEMs was done with
electronics boards based on VFAT chips~\cite{VFAT} 
developed for TOTEM~\cite{TOTEM} 
by INFN Siena-Pisa. The VFAT 
(Very Forward Atlas and Totem) ASIC was designed at CERN using radiation
tolerant technology. It has a 128 channel analog
front-end and produces binary output for each of the channels for tracking. In
addition, it can provide 
a programmable, fast OR function on the input channels depending on the region
of the sensor for triggering. The chip offers adjustable thresholds, gain,
and signal polarity, plus a programmable integration time of the analog input
signals. The signal sampling of the VFAT chip is driven by a 40~MHz internal clock.

During the test beam campaign the readout of all GEM detectors 
with the VFAT
electronics was
digital. The tracker GEMs were read out
in two dimensions with two VFATs connected to the 256 vertical strips, but only one
VFAT connected to 128 out of the 256 horizontal strips. 
The standard
double-mask prototype had a one-dimensional readout with one VFAT connected
to the 128 vertical strips, while
the single-mask and honeycomb triple-GEMs had two-dimensional readout with two
VFATs connected to the 256
vertical strips and one VFAT connected to 128 out of the 256 horizontal
strips. 

\subsection{Data Analysis Results}

The results presented below for the different triple-GEM prototypes were
obtained 
with the data taken during the RD51
SPS test beam campaign from June 28 to July 8, 2010. 

The typical beam profiles for the muon and pion beam as reconstructed with the tracker
GEMs of the RD51 beam telescope are shown in Figure~\ref{fig:beamprofiles}.
For the track reconstruction, events were selected in which the beam telescope GEMs
had only one single cluster of fired strips. Straight tracks were fitted to these
tracker GEM clusters and extrapolated to the detectors under study.  The
alignment of the detectors was done relative to the first tracker GEM, using
the position of the clusters in each detector for single cluster events.

\subsubsection{Standard Double-Mask Triple-GEM}

\begin{figure}[!t]
\centering
\includegraphics[width=3.5in]{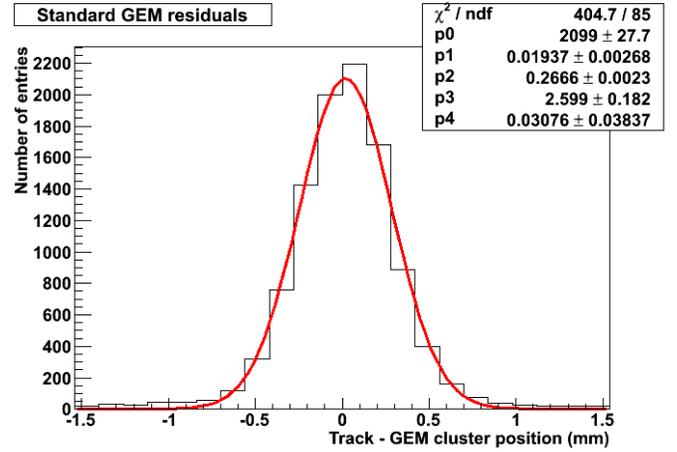}
\caption{The obtained residual distribution for the standard triple-GEM,
  fitted with a Gaussian of the form $p_0\cdot\exp(-0.5\cdot((x-p_1)/p_2)^2)$, plus a first-order polynomial of the form
  $p_3+p_4\cdot x$ to account for 
  noise hits.}
\label{fig:resolutionstandardgem}
\end{figure}

\begin{figure}[!t]
\centering
\includegraphics[width=3.5in]{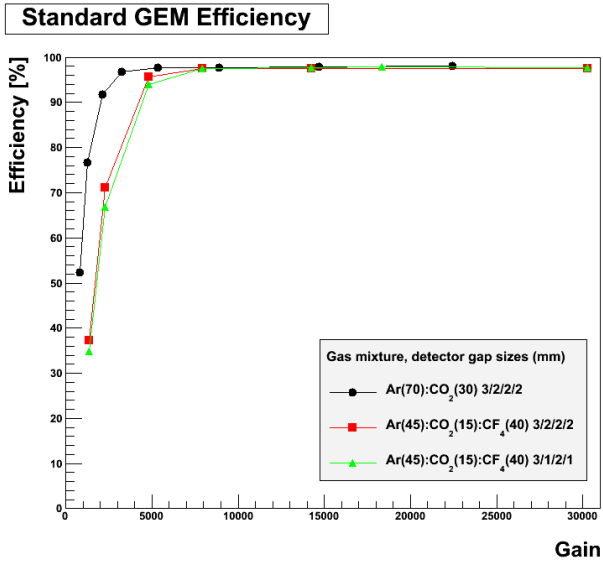}
\caption{Detector efficiency for the standard triple-GEM with different gas
  mixtures and gap size configurations.}
\label{fig:effstandardgem}
\end{figure}

\begin{figure}[!t]
\centering
\includegraphics[width=3.5in]{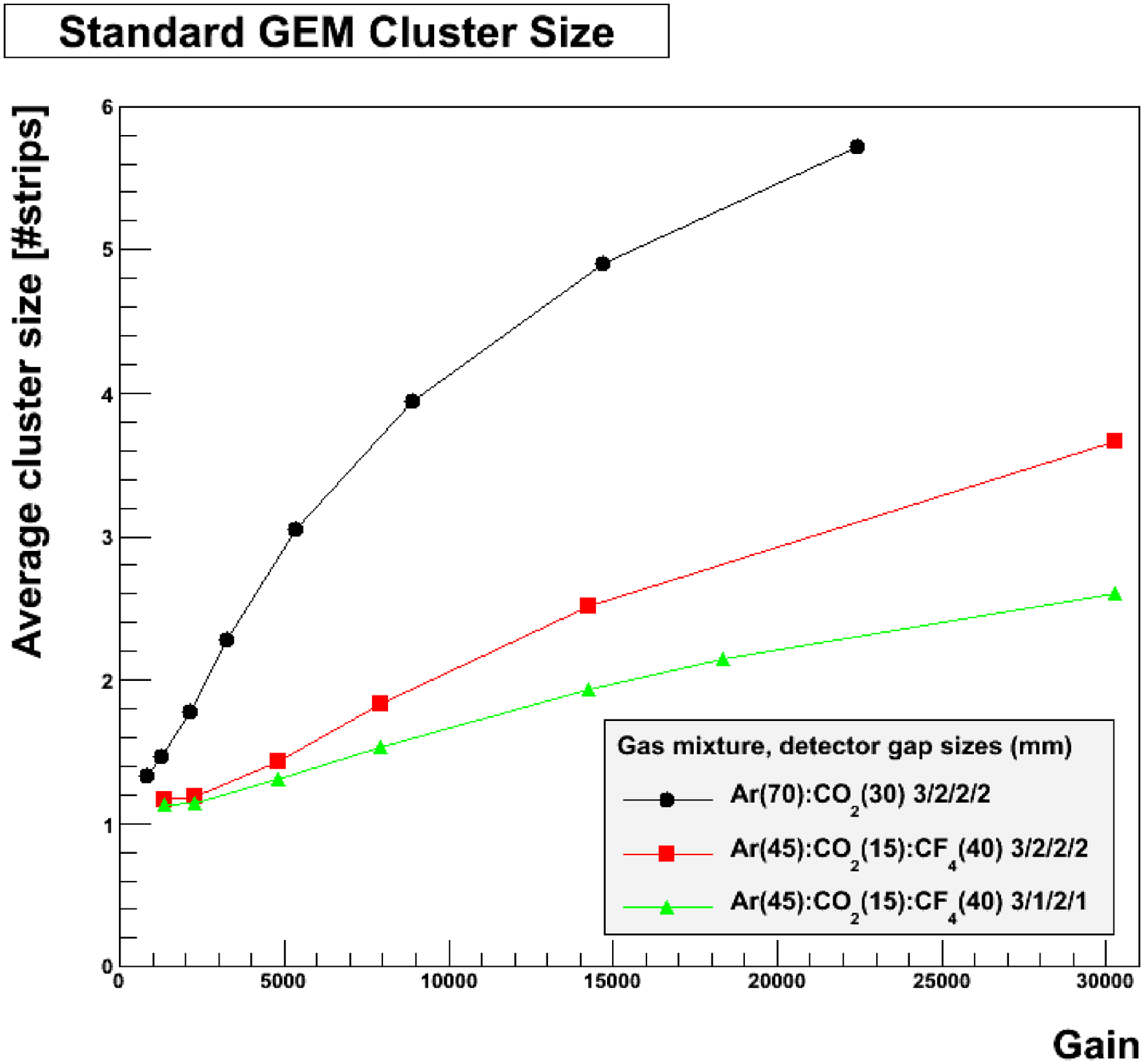}
\caption{Strip cluster size for the standard triple-GEM with different gas
  mixtures and gap size configurations.}
\label{fig:clsizestandardgem}
\end{figure}

\begin{figure}[!t]
\centering
\includegraphics[width=3.5in]{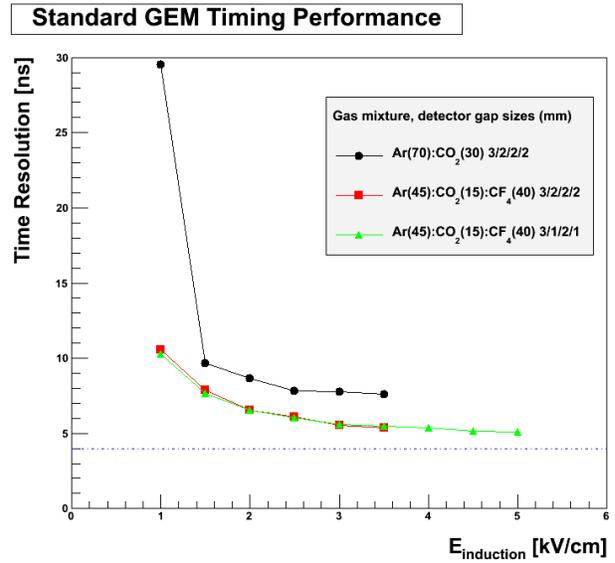}\vspace*{.5cm}
\includegraphics[width=3.5in]{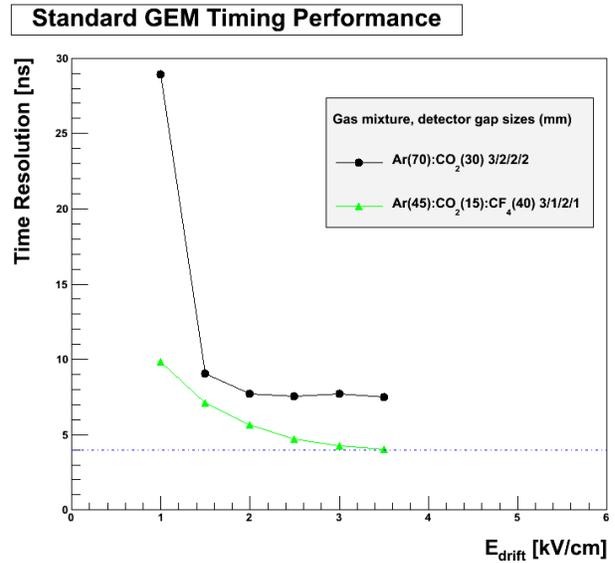}
\caption{Detector timing resolution as function of the induction (top) and
  drift (bottom) field for the standard 
  triple-GEM with 
different gas
  mixtures and gap size configurations.}
\label{fig:timingstandardgem}
\end{figure}

\begin{figure}[!t]
\centering
\includegraphics[width=3.5in]{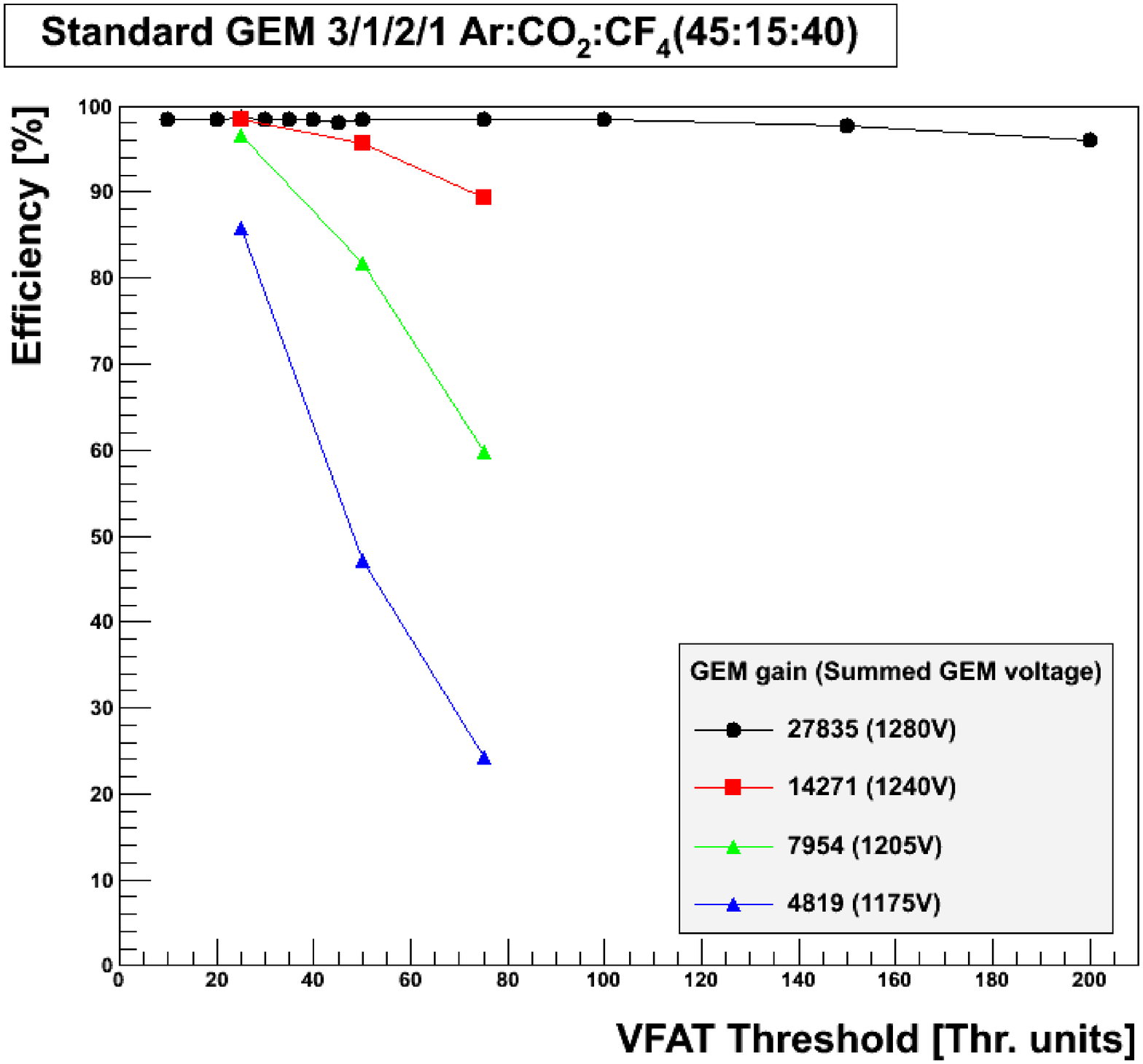}
\caption{VFAT threshold scan for the standard triple-GEM with the 3/1/2/1~mm
  gap size configuration and Ar:CO$_2$:CF$_4$
(45:15:40) gas mixture.}
\label{fig:vfatthrscanstandardgem}
\end{figure}

The typical value obtained for the position resolution for the standard
double-mask triple-GEM operating is about 270~$\mu$m as displayed in 
Figure~\ref{fig:resolutionstandardgem}. This value includes the uncertainty on
the position of the extrapolated track at the detector, and agrees
with the value of 231~$\mu$m (=0.8/$\sqrt{12}$~mm) expected from the 
strip pitch. No strong influence on this resolution value was found from the
detector gap size
configuration, the used gas mixture, or the operating gain.

The measured efficiency for the standard triple-GEM
is displayed in Figure~\ref{fig:effstandardgem}. The efficiency was determined
as function of the detector gain, for different gas mixtures and gap size
configurations. Although for the standard Ar:CO$_2$ (70:30) gas mixture a slightly
better performance is observed for low gain values, 
in each of the cases the efficiency reached the same plateau at
about 98~\% for a gain above 8000. Note also the stability of the detector performance up
to high gains of about $3\cdot 10^4$. 

The effect of the different gas mixtures and the gap size configurations for
the standard triple-GEM on the measured average cluster size, expressed in
number of detector strips, is
shown in Figure~\ref{fig:clsizestandardgem}. Clearly, the use of the Ar:CO$_2$:CF$_4$
(45:15:40) gas mixture yields a much better performance for a digitally read
out detector than the standard Ar:CO$_2$
(70:30) mixture because there are fewer strips per cluster.
Also, the configuration with the smaller transfer 1 and
induction gap size gives
slightly better results.

The timing performance of the standard triple-GEM was studied using a
custom-made high voltage divider that allowed to modify the fields over the
different detector gaps individually. For this study, plastic
scintillators positioned in front and behind the beam telescope were used to
generate a trigger to signal the passage of a beam particle through the detector. 
The spread in arrival time of
the GEM signal from the VFAT board with respect to this external trigger was
measured with a TDC module. In these measurements one has to take into account
the 40~MHz clock cycle of the VFAT chip, which introduced a 25~ns jitter in
the arrival time of the detector signals. 
Note that in case of the LHC, this
jitter can be avoided with a proper synchronization of the VFAT cycle
with the LHC clock.
The obtained time resolution after a deconvolution of the 25~ns VFAT jitter 
is displayed in
Figure~\ref{fig:timingstandardgem}.  
The field across either the
induction or drift gap was varied, while keeping the other fields constant at
(drift, transfer 1, transfer 2, induction gap) 2/3/3/3~kV/cm.
The different gap size
configurations had no visible effect on the timing performance. 
However, the timing performance is clearly better with
the Ar:CO$_2$:CF$_4$ (45:15:40) gas mixture.  With this mixture, a timing
resolution of 4~ns could be obtained.

Based on the observed noise in the
detector without beam, a minimum VFAT threshold of 25 
units\footnote{One VFAT
  threshold unit corresponds to a charge of about 0.08~fC at the input channel
  comparator stage.}
was used for all 
measurements. 
To check the effect of the VFAT threshold on the apparent detector
performance, a VFAT threshold scan was performed for the standard triple-GEM
operating at different gain values as displayed in
Figure~\ref{fig:vfatthrscanstandardgem}. With the VFAT threshold set at 
25 units, no
effect on the efficiency can be observed when the detector is operated at a
gain larger than 10$^4$. 

\subsubsection{Single-Mask Triple-GEM}

\begin{figure}[!t]
\centering
\includegraphics[width=3.5in]{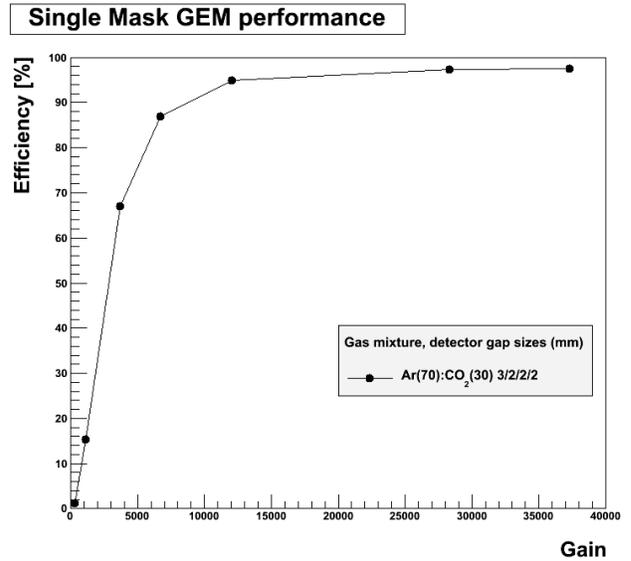}\vspace*{.5cm}
\includegraphics[width=3.5in]{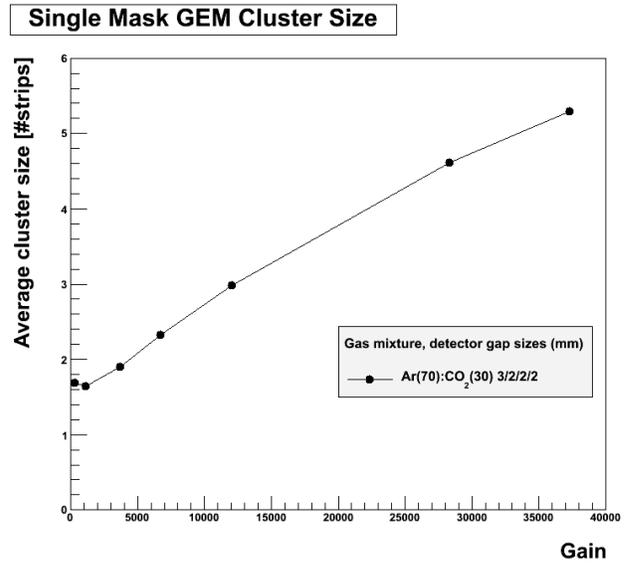}
\caption{Detector efficiency (top) and strip cluster size (bottom) for the single-mask triple-GEM.}
\label{fig:sglmaskgem}
\end{figure}

Several measurements as described above were also performed on the single-mask
triple-GEM to compare its performance to the standard double-mask
triple-GEM. Figure~\ref{fig:sglmaskgem} shows the measured efficiency and
average strip cluster size for the single-mask triple-GEM. The
single-mask GEM reaches a comparable performance level as the corresponding 
double-mask GEM (see Figures~\ref{fig:effstandardgem}-\ref{fig:clsizestandardgem})
albeit the efficiency plateau is attained only at a gain level well above
10$^4$. 

\subsubsection{Honeycomb Triple-GEM}

\begin{figure}[!t]
\centering
\includegraphics[width=3.5in]{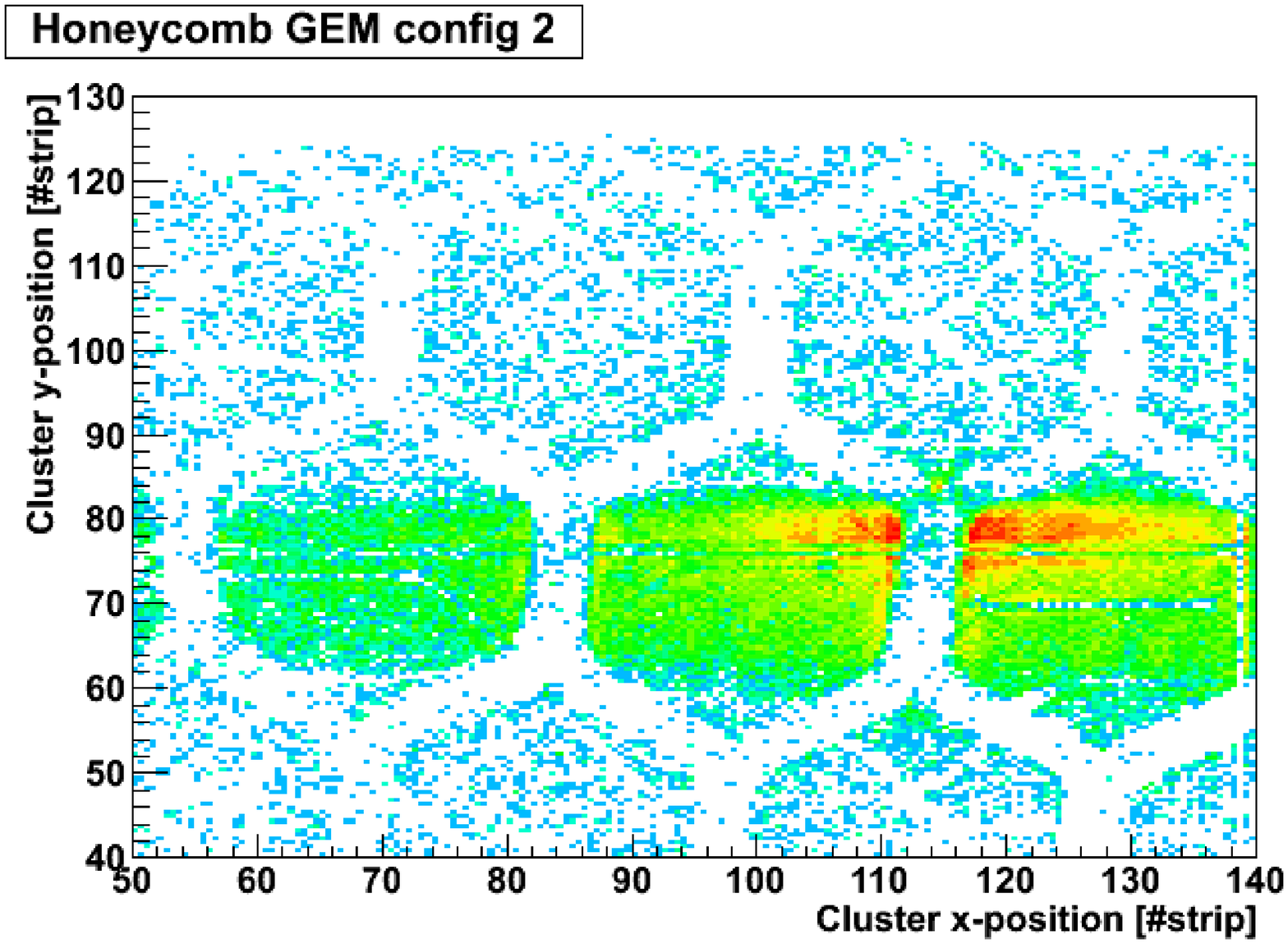}\vspace*{.5cm}
\includegraphics[width=3.5in]{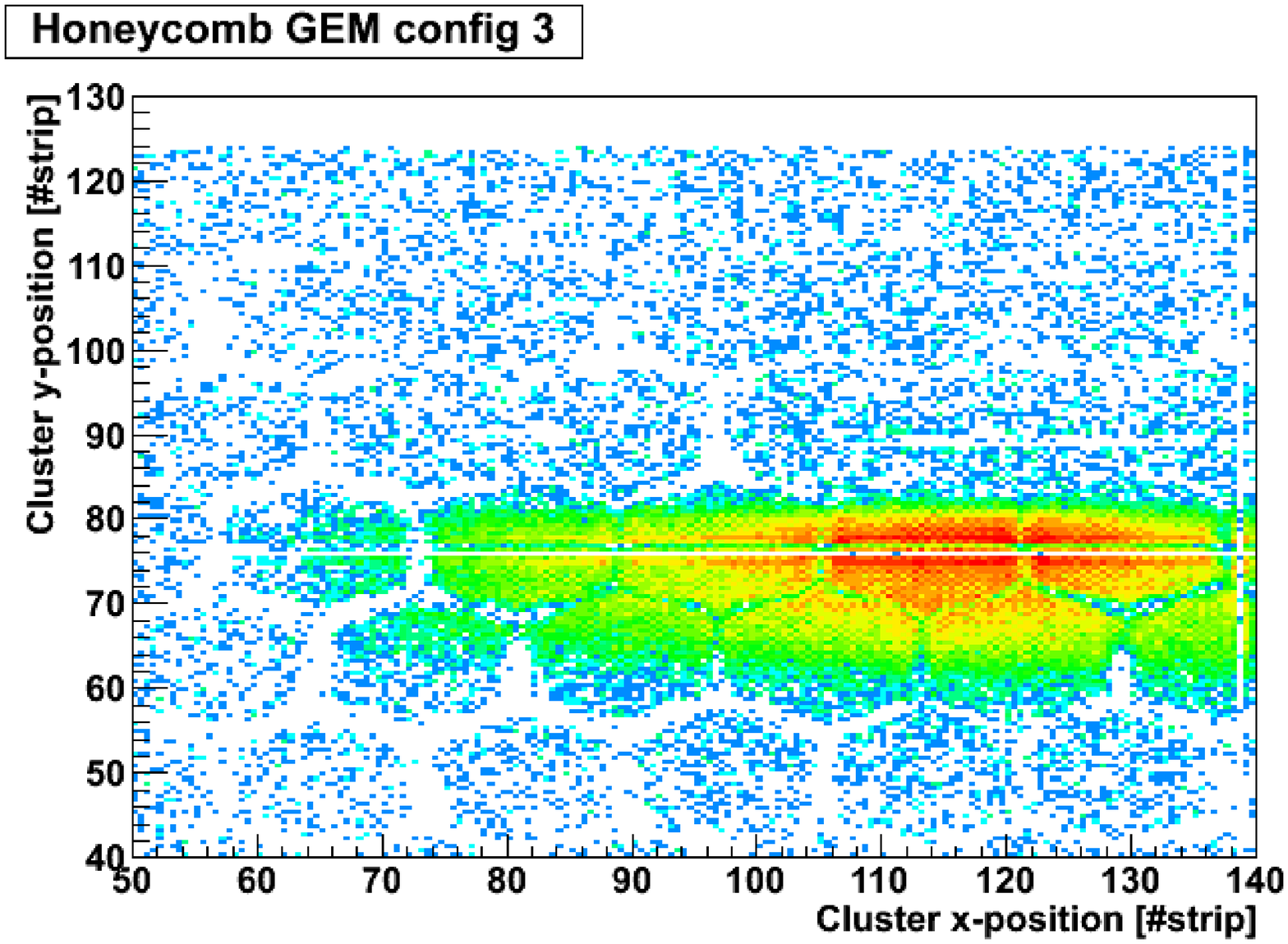}
\caption{Reconstructed cluster positions in the subregions near the beam spot 
of two different honeycomb 
triple-GEM configurations.}
\label{fig:honeycombgems}
\end{figure}

Measurements were performed on the different honeycomb triple-GEMs to check
the effect of the honeycomb spacers on the detector efficiency. For
configuration 1
an overall efficiency of about 50\% was obtained, while configurations 2 and 3
reached an efficiency of about 75\%. The reconstructed cluster positions in
a subregion of the latter two prototypes are displayed in
Figure~\ref{fig:honeycombgems}. One can clearly observe the location of the 
honeycomb spacer material, where a sharp, localized drop in the detector 
efficiency occurs. More
studies
are needed to optimize the geometry of the spacer frames to make this
technique a viable option for detector production.  

\section{Full-Size Triple-GEM Prototype}

\begin{figure}[!t]
\centering
\includegraphics[width=3.5in]{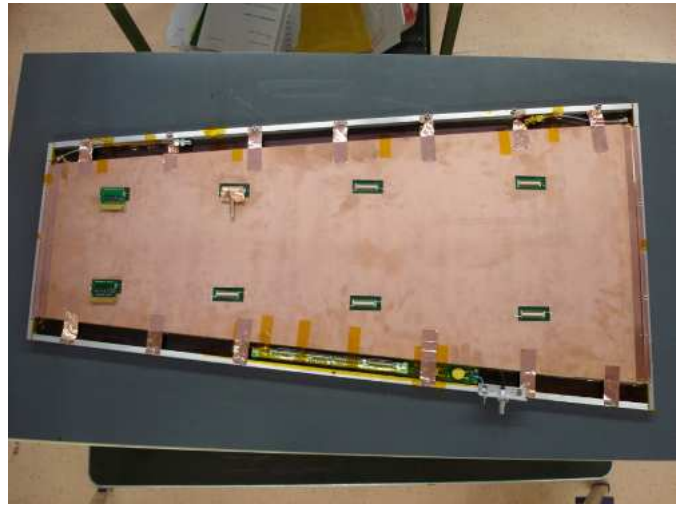}\vspace*{.5cm}
\includegraphics[width=3.5in]{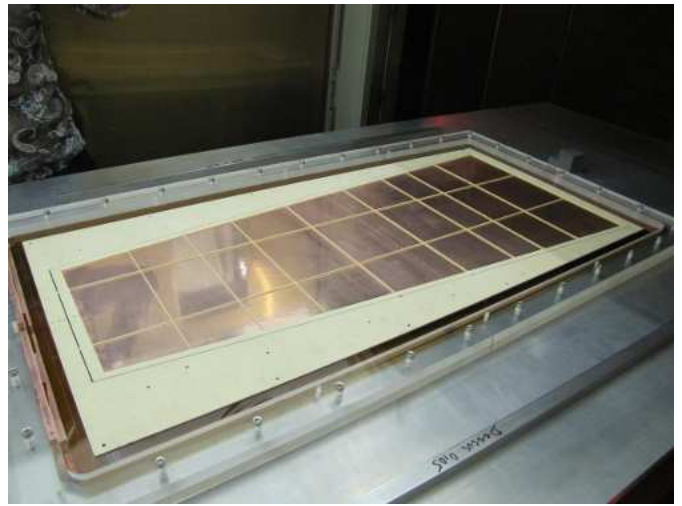}
\caption{The first full-size triple-GEM prototype for CMS. The top picture
  shows the completed detector. The bottom picture
  shows the detector during assembly while glueing the spacer frames.}
\label{fig:largeprototype}
\end{figure}

In October 2010 the construction of a first full-size triple-GEM prototype 
as displayed in Figure~\ref{fig:largeprototype} was completed.
Since it was demonstrated that the single-mask triple-GEM prototype was able
to reach a similar performance level as the double-mask GEM, 
the single-mask technique was used to
produce the GEM foils, as this is a more suitable process to produce large
size 
detectors~\cite{singlemask}.
The detector is divided in 4 $\eta$ partitions containing
256 radial readout strips each oriented along the long side of the detector. Every
$\eta$ partition is read out by two VFAT chips. 
The overall
dimensions of the active area are 990$\times$220-450~mm$^2$
with the strip pitch varying from 0.8 at the small end to 1.6~mm at the large
end of the detector wedge. 
The
GEM foils are sectorized into 35 high voltage sectors transverse to the strip
direction, where each sector has a surface area of about 100~cm$^2$ to limit the
discharge probability. More details on the construction can be found 
in~\cite{largeprototype}.

The detector was tested for the first time 
with a 150 GeV muon/pion beam using the RD51 Tracking GEM telescope at
the CERN SPS H4 beam line. Efficiency scans for different
regions of the detector were
performed with an Ar:CO$_2$ (70:30) gas mixture. 
The data analysis of the measurements is presently
ongoing. 

\section{Summary and Outlook}

Several different 10$\times$10~cm$^2$ triple-GEM prototypes were produced at CERN 
and tested 
at the CERN SPS H4 beamline with a pion/muon beam to study the feasibility 
of using such detector technology for extensions of the CMS muon system.

For the standard double-mask triple-GEM, the best detector performance was 
observed with the use of a Ar:CO$_2$:CF$_4$
(45:15:40) gas mixture instead of Ar-CO$_2$ (70:30) and with a 3/1/2/1~mm gap
size configuration compared to 3/2/2/2~mm. Detector efficiencies up to about
98\% and timing resolutions down to 4~ns were obtained. The detectors could be
operated stably up to high gains of order $3\cdot 10^4$.

The triple-GEM produced with the single-mask technique was seen to reach a
similar performance level as the standard double-mask triple-GEM. 

The use of honeycomb spacer frames in the detector gaps in an attempt to avoid
the need for GEM foil stretching resulted in a clear localized
degradation of the detector efficiency at the position of the frame material. 

A first full-size triple-GEM prototype for the CMS muon system was constructed 
using the
single-mask technique. The detector was tested for the first time with a
muon/pion beam at the CERN SPS H4 beamline.
Lessons learned from the construction and from this first test beam campaign 
will be taken into account during the production of a second, improved full-size
prototype 
in 2011.

\vfill



%

\end{document}